\documentclass[11pt]{article}
\usepackage{newpasp}
\usepackage{psfig}
\usepackage{epsf}
\begin{document}
\title{Results from a Complete \emph{Chandra} Survey of Radio Jets}

\author{D.A. Schwartz (Harvard-Smithsonian Center for Astrophysics), H.L.~Marshall, B.P. Miller (Massachusetts Institute of Technology), \mbox{D.M.~Worrall,}
M. Birkinshaw (University of Bristol), J.E.J. Lovell, D.L.~Jauncey (CSIRO),
\mbox{E.S.~Perlman (U~MD, Baltimore County),} D.W.~Murphy,
R.A. Preston (Jet Propulsion Laboratory)}

\setcounter{page}{111}
\index{Schwartz, D.}
\index{X-ray Jets}


\section{Summary}

We report preliminary results from the first targets observed as part
of a program to image the X-ray jets in a complete sample of
extragalactic radio jets. We have acquired Australian Telescope
Compact Array and Very Large Array data with resolution 1\arcsec\ , matching
the \emph{Chandra} resolution. We tailored our source selection criteria to
generate a list of objects that have continuous jets, or associated
bright knots, with emission extended over scales $>2$\arcsec\ .  We
selected targets from two samples: List A from a VLA sample ($\delta >
0 \deg$) of flat spectrum quasars with core fluxes S$_{5\, \rm{GHz}}> 1$ Jy (Murphy et
al. 1993) and List B from an ATCA survey of flat spectrum Parkes
quasars ($\delta < -20 \deg$) (Lovell 1997), with core flux S$_{2.7\, \rm{GHz}}
\geq 0.34$ Jy. In all, 57 sources comprise the sample, including seven
which were observed as parts of other programs and are not included in
this poster.  In Cycle 3, we were awarded 5 ksec apiece for 20
of the objects, and we report on 16 for which data have been obtained
to date, as shown in Table 1.

\begin{table}[h]
\caption{\emph{Chandra} X-ray Data for the 16 Jets }\label{tabref} 
\scriptsize
\begin{tabular}{lccccccccc}    
\tableline

PKS &   &  &Jet &Jet
&Jet&Jet&  &AGN &List  \\
Name & redshift &
Time(ks) & Counts & Flux$^a$
&Predict$^b$ &Size$^c$&
L$_{\mathrm{jet}}^d$ &
L$_{\mathrm{core}}^d$ & A/B \\

\tableline

 0208-512 &0.999 &5.014 &41 &8.0
& 4.1 &41&1.6 &63  & B\\
 0229+131 &2.059  &5.410&$\le$15 &$\le$2.7 &3.2 &80?&$\le$2.6 &110& B \\

 0413-210  &0.808  &4.862&25 &5.1 &11.4 &29&0.65 &8.8& A \\

 0745+241 &0.410  &5.019&$<$15  &$<$2.9 & 2.9
 && $<$0.09 &5.0  & A \\
 0858-771 &0.49  &4.962&$<$15 &$<$3.0 &3.9 &&$<$0.13 &5.9 & B \\
 0903-573 &0.695  &4.934&18 &3.6 &18.8 &30&0.33 &13 & A\\
 0920-397 &0.591  &4.466&34 &7.5 &10.1 &77&0.48 &7.4 & A\\
 1030-357 &1.455  &5.029&46 &9.0 &4.0 &137&4.1 &35 & B\\
 1046-409 &0.62  &4.330&34 &7.7 &13.5 &32&0.55 &16 & A \\
 1145-676 &    &4.621&$\le$20 &$\le$4.4 &4.5 && &  & B\\

 1202-262& 0.789& 5.074&115&22&11&51&2.7&18 & A\\
 1258-321& 0.01704& 5.426&$<$25&$<$4.5&9.6&&$<$0.0002&0.00056 & A \\

 1424-418 &1.522  &4.475&$<$15 &$<$3.3 &3.2 &&$<$1.7 &110 & B\\
 1655+077 &0.621  &4.825&$<$15 &$<$3.1 &3.2 &&$<$0.22 &12 & B\\
 1655-776 &0.0944  &4.917&$<$15 &$<$3.0 &3.2 &&$<$0.004 &0.050 & A \\
 1828+487 &0.692 &5.329&65 &12.0 &115 &48$^e$&1.1 &47 & A\\

\end{tabular}

\vspace{.07in}

$^{a}${Measured Flux density  at 1 keV, nJy}
$^{b}${Scaled from PKS 0637-753 at 1 keV, nJy}
$^{c}${X-ray size projected on sky, kpc, using H$_{0}$=65, $\Omega_{0}$=0.3,
and $\Omega_{\Lambda}$=0.7}
$^{d}${Rest frame 2 --10 keV luminosity in units of 10$^{44}$ergs s$^{-1}$}
$^{e}${Extended X-ray emission, not specifically from radio jet}
\end{table}

 We detect the jet in X-rays
for seven of these, while for PKS 0229, PKS 1145, and PKS 1828 there
are excess X-ray counts outside the quasar core, but not clearly
associated with the jet. These 10 targets are shown in Figure 1.

Several sources, e.g., PKS 0208-512, PKS 1030-357 and PKS 1202-262,
show strong correlation with the initial radio jet, but with the X-ray
emission declining steeply where the radio jet bends through a large angle. This is
similar to the behavior in PKS 0637-752 (Schwartz et al. 2000). The
radio to X-ray spectral energy distribution (SED) in the case of PKS
0208-512 proves that the emission is not a simple synchrotron
spectrum, due to the low upper limit on the jet optical flux from our Magellan observations (Miller 2002). As in the case of PKS
0637-752, we find that relativistic beaming is necessary to explain
the X-ray emission by inverse Compton scattering on the Cosmic
Microwave Background, if one assumes the radiating electrons are in near
equipartition with the magnetic field in the jet rest frame (Tavecchio et al., 2000; Celotti et al., 2001).

\bigskip


\begin{figure}[t]
\plotone{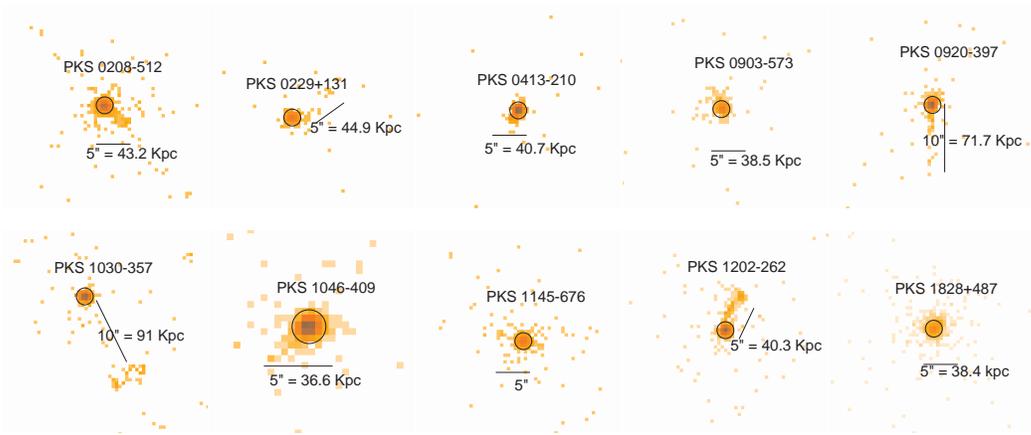} 
\caption{\emph{Chandra} detection of jets and possible extended emission
in our complete survey. The fields are all 31\farcs2 x
29\farcs5. The X-ray data are selected between 0.5 and 7 keV and
binned in 0\farcs492 pixels. Typical background is 0.002 counts per
pixel. The circles are 1\farcs23 radius about the X-ray
centroids, which enclose about 95\% of the flux at 1.5~keV.}
\end{figure}

\vspace{-0.15in}

\acknowledgements This work was supported in part by NASA contract
NAS8-39073 to the \emph{Chandra} X-ray Center, NASA grant GO2-3151C to
SAO, and SAO SV1-61010 to MIT. E.S.P. acknowledges support from NASA
LTSA grant NAG5-9997. This research used the NASA/IPAC Extragalactic
Database (NED) which is operated by the Jet Propulsion Laboratory,
California Institute of Technology, under contract with the National
Aeronautics and Space Administration, and the NASA Astrophysics Data
System Bibliographic Services.


\vspace{-0.15in}

\begin{references}
\reference Celotti, A., Ghisellini, G., \& Chiaberge, M. 2001, \mnras, 321, L1
\reference Lovell, J.E.J. 1997,  Ph.D. Thesis, U. of Tasmania
\reference Miller, B.P. 2002, undergraduate thesis, M.I.T. 
\reference Murphy,  D.W., Browne, I.W.A., \& Perley, R.A. 1993, \mnras,
	264, 298
\reference Schwartz, D. A. et al. 2000, \apj(Letters), 540, L69
\reference Tavecchio, F., Maraschi, L., Sambruna, R. M., \& Urry, C. M.  2000,
\apj(Letters), 544, L23




\end{references}
\end{document}